\documentclass[10pt,aps,twocolumn,prl,showpacs,nofootinbib,showkeys,superscriptaddress,floatfix]{revtex4-1}
\usepackage[dvips]{graphics,graphicx}
\usepackage[colorlinks=true,linktocpage=true,linkcolor=blue,citecolor=blue]{hyperref}
\usepackage[usenames,dvipsnames]{color}
\usepackage{amsmath, amssymb}
\usepackage{multirow}
\usepackage{longtable}
\usepackage{color}
\usepackage{hyperref}
\usepackage{cleveref}
\usepackage[normalem]{ulem}  % \sout{old text} for strikeout

\renewcommand\sout{\bgroup \color{blue} \ULdepth=-.5ex \ULset}
\begin{document}

\preprint{}

\title{Viscosity, non-conformal equation of state and sound velocity in Landau hydrodynamics}

\author{Deeptak Biswas}
\affiliation{Department of Physics, Center for Astroparticle Physics and Space Science, Bose Institute, Kolkata 700091, India}
\author{Kishan Deka}
\affiliation{School of Physical Sciences, National Institute of Science Education and Research, HBNI, Jatni-752050, Odisha, India}
\author{Amaresh Jaiswal}
\affiliation{School of Physical Sciences, National Institute of Science Education and Research, HBNI, Jatni-752050, Odisha, India}
\email{a.jaiswal@niser.ac.in}
\author{Sutanu Roy}
\affiliation{School of Mathematical Sciences, National Institute of Science Education and Research, HBNI, Jatni-752050, Odisha, India}

\date{\today}

%%%%%%%%%%%%%%%%%%%%%%%%%%%%%%%%%%%%%%%%%%%%%%%%%

\begin{abstract}
We find an analytical solution to relativistic viscous hydrodynamics for a 1+1 dimensional Landau flow profile. We consider relativistic Navier-Stokes form of the dissipative hydrodynamic equation, for a non-conformal system with a constant speed of sound, and employ the obtained solution to fit rapidity spectrum of observed pions in $\sqrt{s_{NN}}=$ 200, 17.3, 12.3, 8.76, 7.62, 6.27, 4.29, 3.83, 3.28 and 2.63 GeV collision energies. We find that at the freeze-out hypersurface with improved Landau's freeze-out prescription, the viscous corrections do not affect the rapidity spectra. We demonstrate that the solution of the non-conformal Landau flow lead to a better agreement with the experimental data compared to the conformal ideal solution. We also extract speed of sound from fit to the rapidity spectra for various collision energies and find a monotonous decrease with decreasing collision energies. Appealing to the fact that viscosity has negligible effect on rapidity spectra for Landau's freeze-out scenario, we argue that our calculations provides a framework for extracting the average value of speed of sound in relativistic heavy-ion collisions.
\end{abstract}

%%%%%%%%%%%%%%%%%%%%%%%%%%%%%%%%%%%%%%%%%%%%%%%%%


% 25.75.-q Relativistic heavy-ion collisions
% 24.10.Nz Hydrodynamic models
% 47.75.+f Relativistic fluid dynamics


%%%%%%%%%%%%%%%%%%%%%%%%%%%%%%%%%%%%%%%%%%%%%%%%%

\maketitle

%%%%%%%%%%%%%%%%%%%%%%%%%%%%%Introduction%%%%%%%%%%%%%%%%%%%%%%%%%%%%%%%%%

\emph{Introduction}: 
Heavy ion collisions at relativistic energies offer the possibility to create strongly interacting hot and dense matter over extended region \cite{Lee:1974ma, Collins:1974ky, Itoh:1970uw}. The space-time evolution of this hot and dense matter, created at Relativistic Heavy ion Collider (RHIC) and the Large Hadron Collider (LHC), has been successfully modeled using relativistic dissipative hydrodynamic simulations \cite{Romatschke:2007mq, Luzum:2008cw, Luzum:2009sb, Song:2010mg, Luzum:2010ag, Schenke:2011tv, Gale:2012rq, Bhalerao:2015iya, Jaiswal:2016hex}. However, the first hydrodynamic approach to describe nucleus-nucleus collisions was proposed by Landau in 1953 where he studied longitudinal expansion along the collision axis \cite{Landau:1953gs}. In his approach, Landau considered a non-dissipative expansion of the medium and obtained an approximate analytical solution of ideal hydrodynamic equations (commonly refered to as the Landau hydrodynamics) which led to Gaussian like rapidity distributions of produced particles \cite{Landau:1953gs}. Later a plateau like rapidity distribution was proposed in Hwa-Bjorken hydrodynamics which considered a boost-invariant framework \cite{Hwa:1974gn, Bjorken:1982qr}. However, the boost-invariance symmetry was only applicable in the mid-rapidity region of ultra-relativistic heavy ion collisions. On the other hand, the transverse-momentum ($p_T$) integrated yield over the whole rapidity region has an overall better agreement with the Gaussian structure suggested by Landau \cite{Murray:2004gh, Bearden:2004yx, Murray:2007cy, Steinberg:2004wx, Steinberg:2007iv, Wong:2008ex}.

Several problems in high-energy heavy ion collisions, such as heavy quark propagation and the interaction of jets or quarkonia with the hot and dense matter, requires a realistic but simple description of the evolution of the produced medium. Successes of Landau hydrodynamics in explaining the total charged multiplicities, rapidity distribution and limiting fragmentation \cite{Murray:2004gh, Bearden:2004yx, Murray:2007cy, Steinberg:2004wx, Steinberg:2007iv, Wong:2008ex} indicate that it can provide such a framework. Landau hydrodynamics is based on the assumption that the initial stages of relativistic heavy ion collisions experiences a fast longitudinal expansion accompanied by a slower expansion in the transeverse plane. One can obtain analytical solution for the one-dimensional longitudinal expansion \cite{Landau:1953gs} which has been analyzed in great detail in the literature \cite{Belenkij:1956cd, Khalatnikov:1954, Belenkij:1956, Rosental:1957, Milekhin:1958a, Milekhin:1958b, Amai:1957, Carruthers:1973ws, Cooper:1974mv, Cooper:1974qi, Chadha:1974qs, Srivastava:1992xb, Srivastava:1992cg, Srivastava:1992gh, Mohanty:2003va, Hama:2004rr, Aguiar:2000hw, Pratt:2008jj, Bialas:2007iu, Csorgo:2006ax, Beuf:2008vd, Osada:2008cn}. After sufficiently large transverse expansion, i.e., when the magnitude of the transverse displacement becomes larger than the initial transverse dimension, the fluid element expands with a frozen rapidity. The final rapidity distribution of particles is then given by that at the freeze-out time which leads to good agreement with the experimental data \cite{Murray:2004gh, Bearden:2004yx, Murray:2007cy, Steinberg:2004wx, Steinberg:2007iv, Wong:2008ex, Srivastava:1992xb, Srivastava:1992cg, Srivastava:1992gh, Mohanty:2003va, Hama:2004rr, Aguiar:2000hw, Pratt:2008jj, Bialas:2007iu, Csorgo:2006ax, Beuf:2008vd, Osada:2008cn}.

Despite the fact that the gross features of many measured quantities are reproduced well, the Landau hydrodynamics can, at best, be considered a good first approximation. One can expect corrections and improvements based on physical considerations and the requirement to explain experimental data. The need to improve up on Landau's original rapidity distribution should not come as a surprise, as the original Landau distribution was intended to be qualitative. On the other hand a more quantitative approach in order to explain experimental results and hence extract additional information about the properties of the QCD matter formed in relativistic heavy ion collisions is desirable. One such important improvement to Landau hydrodynanics would be to include dissipation in the evolution which will provide a platform to analytically estimate the transport coefficient of the QCD matter by analyzing the rapidity spectra. While there has been attempts to improve the Landau model by including viscous corrections analytically \cite{Hoang:1973uz, Chaichian:1976nt, Hoang:1977wf}, the resultant solutions were not conclusive and/or did not show the correct trend towards explaining the observed experimental data.

In this article, we consider the relativistic Navier-Stokes equation for viscous evolution of the hot and dense matter formed in high energy heavy-ion collisions. We find analytical solution of the viscous evolution equation, for non-conformal system having constant speed of sound, with 1+1 dimensional Landau flow profile. We find that for Landau's prescription of freeze-out scenario, the viscous effects do not effect the rapidity spectrum of the produced particles. We employ the obtained solution to fit rapidity spectrum of observed pions in $\sqrt{s_{NN}}=$ 200, 17.3, 12.3, 8.76, 7.62, 6.27, 4.29, 3.83, 3.28 and 2.63 GeV collision energies. We show that the obtained solutions of Landau flow with non-conformal equation of state leads to a better agreement with the experimental data compared to the conformal Landau flow solution. We also find that the value of speed of sound, obtained by fitting the experimental results, show monotonic decrease with decreasing collision energies. Appealing to the fact that viscosity has negligible effect on rapidity spectra for Landau's freeze-out scenario, we advocate that our calculations provide a framework for extracting the average value of sound velocity of QCD medium produced in heavy-ion collisions.

%%%%%%%%%%%%%%%%%%%%%%%%%%%%%Model%%%%%%%%%%%%%%%%%%%%%%%%%%%%%%%%%%%%%%%%

\emph{Viscous Landau flow}: 
Hydrodynamic equations follow from the principle of conservation of energy and momentum of a continuous medium, which leads to vanishing of four-divergence of the energy-momentum tensor for a relativistic system. In absence of any conserved charges, the energy-momentum tensor of a relativistic fluid, with dissipative terms from Navier-Stokes theory, can be written as \cite{Landau_book}
\begin{equation}\label{Tmunu}
T^{\mu \nu}=\epsilon\, u^\mu u^\nu - (P-\zeta\theta)\, \Delta^{\mu \nu} + 2\eta\sigma^{\mu\nu},
\end{equation}
where $\epsilon$ is the local energy density, $P$ is the thermodynamic pressure, $u^\mu$ is the fluid four-velocity and, $\eta$ and $\zeta$ are the coefficients of shear and bulk viscosity, respectively. We also define a projection operator $\Delta^{\mu\nu}\equiv g^{\mu\nu}-u^\mu u^\nu$ which is orthogonal to $u^\mu$, four-divergence of fluid velocity $\theta\equiv\partial_\mu u^\mu$ and a derivative operator $\nabla^\mu\equiv\Delta^{\mu\nu}\partial_\nu$ which is also orthogonal to $u^\mu$. We follow mostly minus metric convention, i.e., $g^{\mu\nu}={\rm diag}(1,-1,-1,-1)$. Using these definitions, the shear tensor can be written as $\sigma^{\mu\nu}\equiv\frac{1}{2}\left(\nabla^\mu u^\nu + \nabla^\nu u^\mu \right)-\frac{1}{3}\Delta^{\mu\nu}\nabla_\alpha u^\alpha$. In this work, we shall use the non-conformal equation of state, $P=c_s^2\epsilon$, where the speed of sound $c_s^2$ will be assumed to be constant for simplicity.

Following the seminal work of Landau \cite{Landau:1953gs}, we consider the longitudinal dynamics in the collision of two identical nuclei moving along the $z$-direction. For longitudinal expansion, the hydrodynamic equation, $\partial_\mu T^{\mu\nu}=0$, leads to \cite{Landau:1953gs, Wong:2008ex}
\begin{equation}\label{tz}
\frac{\partial T^{00}}{\partial t} + \frac{\partial T^{03}}{\partial z} = 0,\qquad
\frac{\partial T^{03}}{\partial t} + \frac{\partial T^{33}}{\partial z} = 0,
\end{equation}
where, we have used the notation $(t,x,y,z)\equiv(x^0,x^1,x^2,x^3)$ for co-ordinate labels. For one-dimensional expansion along $z$-axis, we introduce the longitudinal fluid rapidity, $y$, in terms of which we can represent the non-zero velocity fields $u^0=\cosh{y}$ and $u^3=\sinh{y}$. In terms of the light-cone variables, $t_\pm\equiv t\pm z$, Eq.~(\ref{tz}) can be written as
\begin{align}
\frac{\partial}{\partial t_+}\left[c_+\epsilon - \xi\,\nabla u \right]e^{2y} + \frac{\partial}{\partial t_-}\left[c_-\epsilon + \xi\,\nabla u \right] &= 0, \label{tptm1}\\
\frac{\partial}{\partial t_+}\left[c_-\epsilon + \xi\,\nabla u \right] + \frac{\partial}{\partial t_-}\left[c_+\epsilon - \xi\,\nabla u \right]e^{-2y} &= 0, \label{tptm2}
\end{align}
where $c_\pm\equiv1\pm c_s^2$ and $\nabla u\equiv\partial u^0/\partial t +\partial u^3/\partial z =e^y\,\partial y/\partial t_+- e^{-y}\,\partial y/\partial t_-$ is defined to simplify notations and $\xi\equiv\zeta+4\eta/3$. In this article, we solve the above set of differential equations for evolution of energy density assuming the flow profile to be given by ideal Landau hydrodynamics.

For a flow which is not boost-invariant, the fluid rapidity $y$ is not the same as the space-time rapidity $\eta_s\equiv\frac{1}{2}\ln(t_+/t_-)$. The notion of Landau flow is based on the assumption that the difference between the fluid rapidity and space-time rapidity is small. Therefore one can express the fluid rapidity in terms of the space-time rapidity as \cite{Landau:1953gs}
\begin{equation}\label{lnd_fl}
e^{2y}=f\,e^{2\eta_s}=f\,\frac{t_+}{t_-},
\end{equation}
where $f$ is a slowly varying (logarithmic) function of $t_+$ and $t_-$, of order unity, such that the derivatives of $f$ and quadratic and higher-order terms in $\log f$ could be neglected. Keeping in mind that $\sqrt{f}+1/\sqrt{f}\simeq 2$, we find that $\nabla u=1/\sqrt{t_+t_-}$ for the expansion profile given in above equation. We further introduce another change of variables, $y_\pm\equiv\ln\left(t_\pm/\Delta\right)$, where $\Delta=2R/\gamma$ is the longitudinally Lorentz contracted diameter of each colliding nuclei. In terms of these new variables, Landau obtained $f=\sqrt{y_+/y_-}$ for ideal hydrodynamic evolution, which is indeed a logarithmic function of $t_+$ and $t_-$ of order unity \cite{Landau:1953gs}. With this velocity profile, the solution for evolution of energy density in case of ideal hydrodynamics, i.e., $\xi=0$, is given by
\begin{equation}\label{sol_id}
\epsilon_{id}=\epsilon_0\exp\!\left[ -\frac{c_+^2}{4\,c_s^2}\left( y_+ + y_-\right) + \frac{c_+c_-}{2\,c_s^2}\sqrt{y_+\, y_-}  \right]\!,
\end{equation}
where $\epsilon_0$ is related to the initial energy density. It is important to note that, in the conformal limit, we reproduce the original results of Landau \cite{Landau:1953gs, Wong:2008ex}. 

In the following, we assume the flow profile to be given by ideal Landau hydrodynamics and find a solution for the evolution of energy density. Changing evolution variables to $y_\pm$, we see that Eq.~(\ref{tptm1})~$+$~Eq.~(\ref{tptm2}) leads to
\begin{equation}\label{ypym1}
f\,\frac{\partial\epsilon}{\partial y_+} + \frac{\partial\epsilon}{\partial y_-} + \frac{1+f}{2}\left[ c_+\epsilon - \frac{\xi}{\Delta}\,e^{-(y_+ + y_-)/2} \right] = 0.
\end{equation}
We can generate another linearly independent equation from Eq.~(\ref{tptm1})~and~(\ref{tptm2}) by considering the combination Eq.~(\ref{tptm1})~$-$~Eq.~(\ref{tptm2}),
\begin{align}\label{ypym2}
f\frac{\partial\epsilon}{\partial y_+} - \frac{\partial\epsilon}{\partial y_-} + \frac{(f-1)c_+}{2\,c_s^2}\epsilon  
&- \frac{1}{c_s^2\Delta}\left( f\frac{\partial\xi}{\partial y_+} - \frac{\partial\xi}{\partial y_-}\right)\nonumber\\
&\times e^{-(y_+ + y_-)/2} = 0.
\end{align}
It is important to note that for collision of two identical nuclei, the reflection symmetry about the transverse plane in the centre-of-mass frame must lead to evolution equations invariant under $y_+\leftrightarrow y_-$ interchange. Therefore the hydrodynamic equations describing the evolution of the matter formed in these collisions should be even under parity transformation. Keeping in mind that we assume Landau flow profile, i.e., Eq.~(\ref{lnd_fl}) with $f=\sqrt{y_+/y_-}$, it is easy to see that the left hand side of Eq.~(\ref{ypym1}) has even parity whereas that of Eq.~(\ref{ypym2}) has odd parity. Therefore, the solution of Eq.~(\ref{ypym1}) should lead to the evolution of energy density for viscous Landau flow in symmetric nucleus-nucleus collisions.

To make progress, we assume the ratio $\xi/s$ to be a constant where $s$ is the entropy density. While this is a valid assumption in conformal case, it is not be strictly true for a non-conformal system. Therefore, for the case of constant $\xi/s$, one can write $\xi=\alpha\,\epsilon^{1/(1+c_s^2)}=\alpha\,\epsilon^{1/c_+}$, where $\alpha$ is a constant. Substituting in Eq.~\eqref{ypym1} and rearranging, we get 
\begin{equation}\label{ypym1_conf}
f\frac{\partial\epsilon}{\partial y_+} + \frac{\partial\epsilon}{\partial y_-} = \frac{1+f}{2}\!\left[ \frac{\alpha}{\Delta}\epsilon^{\frac{1}{c_+}}e^{-\frac{1}{2}(y_+ + y_-)} - c_+\epsilon \right]\!.
\end{equation}
Using method of characteristics, we get,
\begin{equation}\label{meth_char}
\frac{dy_+}{f} = \frac{dy_-}{1} = \frac{2\,d\epsilon}{(1+f)\!\left[ \frac{\alpha}{\Delta}\epsilon^{\frac{1}{c_+}}e^{-\frac{1}{2}(y_+ + y_-)} - c_+\epsilon \right]}.
\end{equation}
The above equations can be solved analytically to obtain
\begin{equation}\label{sol_vis}
\epsilon = \left[ \mathcal{F}\!\left(\! \frac{y_+}{f}\!-\!y_- \!\!\right)\! e^{-\frac{1}{2}(1+f)c_s^2y_-} - \frac{c_s^2\alpha}{c_+c_-\Delta}e^{-\frac{1}{2}(y_++y_-)} \right]^{\!\frac{c_+}{c_s^2}},
\end{equation}
where, $\mathcal{F}$ is an arbitrary function of the argument given in parenthesis and can be determined by comparing the above solution with the ideal solution given in Eq.~\eqref{sol_id}, in the limit of vanishing viscosity, i.e., $\alpha=0$. We get
\begin{equation}\label{id_vis_comp}
\mathcal{F}\!\left( \frac{y_+}{f}-y_- \right) = \epsilon_0^{c_s^2/c_+} \exp\! \left[ \frac{c_- - c_+ f}{4}\left(\frac{y_+}{f}-y_- \right) \right].
\end{equation}
Substituting the above value of $\mathcal{F}$ in Eq.~\eqref{sol_vis}, we obtain the final solution for energy density evolution with Landau flow profile
\begin{equation}\label{sol_vis_fin}
\epsilon=\left[ g(\alpha)\,\epsilon_{id}^{c_s^2/c_+} -\frac{c_s^2\,\alpha}{c_+c_-\Delta}\,e^{-(y_+ + y_-)/2} \right]^{\frac{c_+}{c_s^2}},
\end{equation}
where $g(\alpha)$ is an arbitrary function of $\alpha$ such that $g(0)=1$. It is easy to see that the above form of energy density indeed satisfy Eq.~\eqref{ypym1}.  

We note that while the above solution is formally similar to that obtained in Ref.~\cite{Chaichian:1976nt}, there is still some leftover freedom in our solution given in Eq.~\eqref{sol_vis_fin} for the functional form of $g(\alpha)$. One way to fix this residual freedom is by considering the longitudinal boost-invariant Bjorken limit of Eq.~\eqref{sol_vis_fin} and comparing with the corresponding solution of viscous hydrodynamics in Bjorken case \cite{Hosoya:1983xm, Kouno:1989ps}. Doing this exercise, we obtain
\begin{equation}\label{glapha}
g(\alpha)= 1 + \frac{\alpha\, c_s^2}{c_+\,c_- \,\epsilon_0^{c_+/c_s^2}\,\Delta},
\end{equation}
where $\epsilon_0$ is the energy density corresponding to the initial proper time $\tau_0=\Delta$. Equations \eqref{sol_vis_fin} and \eqref{glapha} together constitute the analytical solution of viscous Landau hydrodynamics and represents the main results of the present work which will be used subsequently for calculating the rapidity spectra of produced hadrons in heavy-ion collisions. Nevertheless, as demonstrated in the following, we find that the effect of $\alpha$ (i.e., viscosity) on rapidity spectra of produced particles is negligible when we consider Landau's prescription for freeze-out hypersurface.

%%%%%%%%%%%%%%%%%%Results/Discussions%%%%%%%%%%%%%%%%%%%%%%%%%%%%%%%%%%%%%

\emph{Rapidity Distribution}:
Following Landau's arguments \cite{Landau:1953gs}, we assume that the one-dimensional solution is applicable until the expansion of the fluid element in the transverse direction is of the same order as the transverse dimension of the system. We find that if one considers the longitudinal and transeverse expansion independently, having a slower expansion with constant acceleration in the transverse direction \cite{Landau:1953gs, Wong:2008ex}, then the transverse expansion does not get any correction due to viscosity. Under this approximation the freeze-out time also remains the same as ideal case \cite{Wong:2008ex}. While there are many possible freeze-out conditions, the successes of Landau hydrodynamics suggests that Landau's freeze-out criteria can be a good first approximation and the correction is likely to be small and scale with the Landau freeze-out proper time. 

Following Wong's modified prescription \cite{Wong:2008ex} of Landau's freeze-out criteria \cite{Landau:1953gs}, we consider the transverse expansion using the non-conformal equation of state. In this case, one obtains a freeze out hypersurface where the freeze out time is given by, 
\begin{equation}\label{tfo}
t_{\rm FO} = 2R\sqrt{\frac{1+c_s^2}{c_s^2}}\,\cosh y.
\end{equation}
With the help of above expression rapidity variables at the freeze-out hypersurface takes the form $y_\pm=y_b'\pm y$ where $y_b'\equiv\frac{1}{2}ln[c_+/(4c_s^2)] + y_b$ and $y_b\equiv\ln(\sqrt{s_{NN}}/m_p)$ is the beam rapidity with $m_p$ being mass of the proton \cite{Wong:2008ex}. Keeping in mind that the beam rapidity is the largest rapidity achievable by a fluid element, we see that the term proportional to $\alpha$ in Eq.~\eqref{sol_vis_fin} becomes negligible at freeze-out and could be ignored in the first approximation. Moreover, keeping in mind that,
\begin{equation}\label{taufo}
e^{-(y_++y_-)/2}=\frac{\Delta}{\tau_{\rm FO}}=\frac{1}{\gamma}\sqrt{\frac{1+c_s^2}{c_s^2}},
\end{equation}
where $\tau_{\rm FO}$ is the proper-time at freeze-out, one can also see that this term in Eq.~\eqref{sol_vis_fin} is negligible because of the large Lorentz factor, $\gamma$, of the colliding nuclei. From the above equation, we also see that the freeze-out hypersurface has constant proper-time.

In order to understand the physical implications of this freeze-out scenario, we first note that viscosity indeed affects particle productions. This is well known from viscous hydrodynamic simulations of heavy-ion collisions where it is observed that inclusion of viscosity leads to change in the transverse momentum spectra \cite{Teaney:2003kp, Song:2009rh}. However, our claim is that although viscosity influences particle production via the overall normalization factor $g(\alpha)$, the shape of the rapidity distribution is not affected significantly. In this scenario, the bulk of observed particles come from central hotter region where the viscosity does not play significant role. Viscosity plays important role at the edges of the fireball where the gradients are large but the temperature is small, leading to negligible contribution in the rapidity spectra. 

It is important to note that for ideal evolution, the ratio of entropy density to number density, $s/n$, is a conserved quantity. This is not expected to hold when one has dissipation in the system. On the other hand, we also observe that the entropy density does not get any direct correction from dissipative term in the relativistic Navier-Stokes equation, i.e., $s\sim \epsilon^{1/c_+}$ and hence $s/n$ is approximately conserved for viscous evolution. Moreover, in the present work, we saw that the viscous correction to energy density evolution can be ignored at freeze-out, and therefore the final expression for rapidity distribution turns out to be proportional to entropy density which is given by,
\begin{equation}\label{rap_sp}
\dfrac{dN}{dy} \sim \exp\!\left( \frac{c_-}{2c_s^2}\,\sqrt{{y_b'}^2 - y^2} \right).
\end{equation}
We see that by setting $c_s^2=1/3$ in the above equation, the ideal rapidity spectrum for conformal system is recovered \cite{Landau:1953gs, Wong:2008ex}. We use the above expression of rapidity spectrum to fit the observed pion spectra in heavy-ion collision experiments at available centre-of-mass energies.  

\begin{figure}[t]
\includegraphics[width=\linewidth]{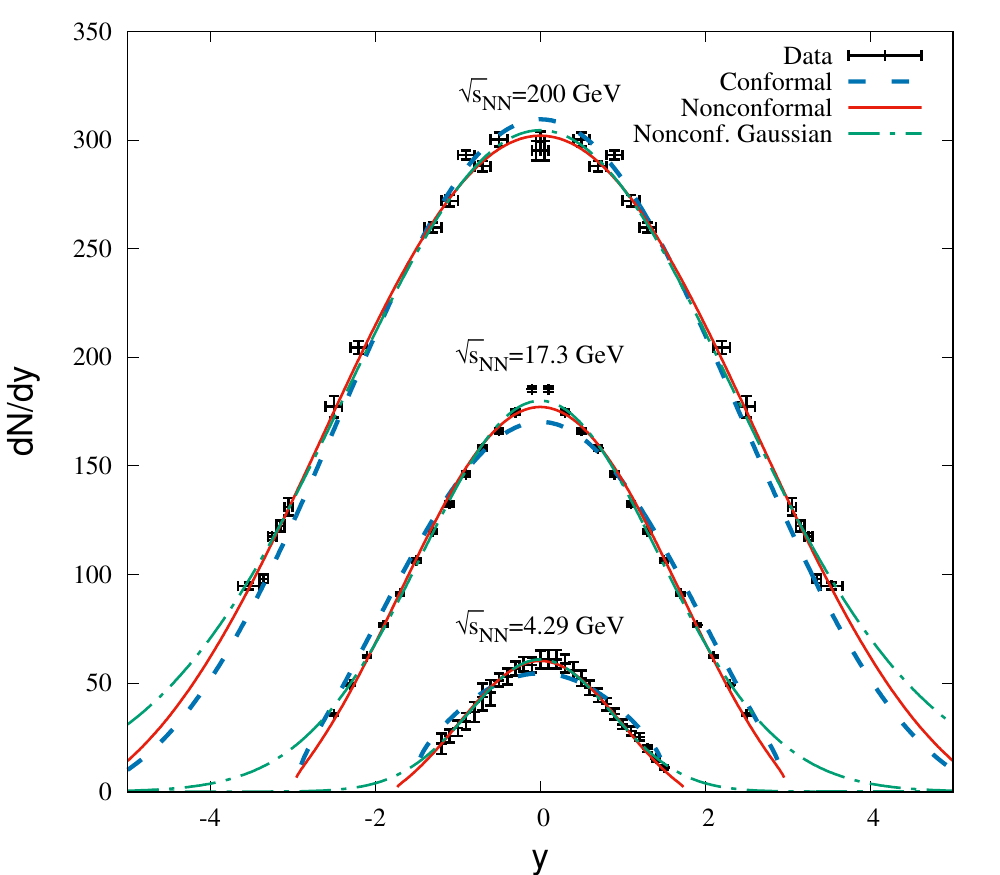}
\vspace*{-0.3cm} 
\caption{(Color online) Rapidity spectrum of pions fitted using Eq.~\eqref{rap_sp} (red solid curves) and Eq.~\eqref{rap_gaus} (green dashed-dotted curves), for three representative collision energies: $\sqrt{s_{NN}}=$ 200, 17.3 and 4.29 GeV. Also shown are the fit result using conformal solution of Landau hydrodynamics \cite{Wong:2008ex} (blue dashed curves) and the experimental results (black symbols). Experimental data are from Refs.~\cite{Klay:2003zf, Alt:2007aa, Afanasiev:2002mx, Bearden:2004yx}. }
\label{spectra}
\end{figure}

In Fig.~\ref{spectra}, we show rapidity spectrum of pions fitted using Eq.~\eqref{rap_sp} for three representative collision energies, $\sqrt{s_{NN}}=$ 200, 17.3 and 4.29 GeV (red solid curves) compared with the fit result using conformal solution of Landau hydrodynamics \cite{Wong:2008ex} (blue dashed curves) and the experimental results (black symbols). From the figure, we see that a better fit is obtained using the non-conformal solutions for these collision energies. We have also fitted the rapidity spectrum of pions for $\sqrt{s_{NN}}=$ 12.3, 8.76, 7.62, 6.27, 3.83, 3.28 and 2.63 GeV and found that there is an overall better fit with solutions from non-conformal equation of state.

\begin{figure}[t]
\includegraphics[width=\linewidth]{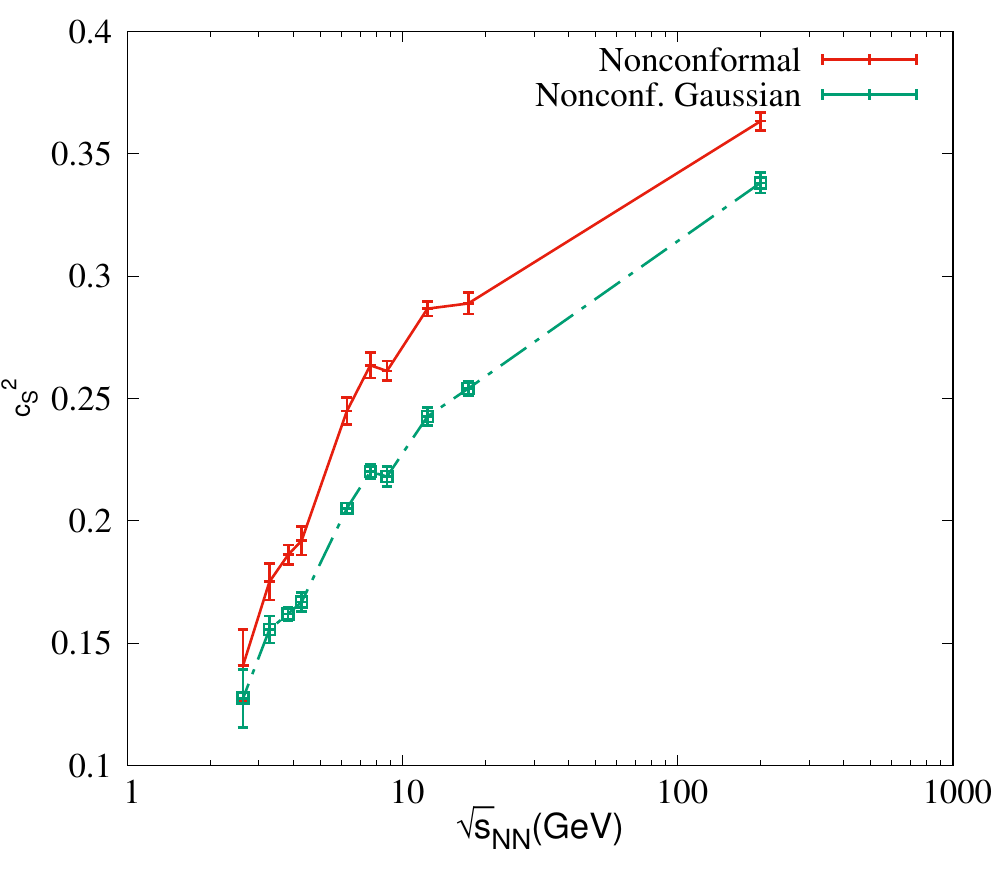}
\vspace*{-0.3cm} 
\caption{(Color online) Squared speed of sound, extracted by fitting the pion rapidity spectra using the rapidity distribution obtained using non-conformal solution given in Eq.~\eqref{rap_sp} (red solid line) and also using the Gaussian distribution given in Eq.~\eqref{rap_gaus} (green dashed-dotted line), over various collision energies. The error bars corresponds to standard error from least-square fit of the fit parameters.}
\label{cs2}
\end{figure}

In Fig.~\ref{cs2}, we show a plot of squared speed of sound, extracted by fitting the pion rapidity spectra using Eq.~\eqref{rap_sp}, over various collision energies (red solid line). The error bars in the plot corresponds to standard error from least-square fit of the fit parameters. We see that at $\sqrt{s_{NN}}=$ 200 GeV, the fitted value of $c_s^2$ is slightly larger than $1/3$ which has also been observed in Refs.~\cite{Gazdzicki:2010iv, Gao:2015mha}. With decrease in collision energy, the extracted value of $c_s^2$ is seen to decrease, which is in agreement with lattice QCD predictions for temperature dependence of $c_s^2$ \cite{Bazavov:2014pvz, Ding:2014kva}. However, with lower collision energies, we do not find a minimum in the $\sqrt{s_{NN}}$ dependence of $c_s^2$ which is in contrast with earlier calculations of Ref.~\cite{Gazdzicki:2010iv}. This may be due to the fact that the rapidity distribution given in Eq.~\eqref{rap_sp}, which is used to fit the data, and the expression of $y_b'$ is different in the present work and in Ref.~\cite{Gazdzicki:2010iv}, as explained below. 

Based on our analysis of the rapidity spectra, we found that the width of the Gaussian profile is controlled by $c_s^2$ and hence the equation of state of the medium. This is easy to see from Eq.~\eqref{rap_sp}. For rapidities small compared to the beam rapidity, one can rewrite Eq.~\eqref{rap_sp} to obtain the well known Gaussian rapidity distribution
\begin{equation}\label{rap_gaus}
\dfrac{dN}{dy} \sim \exp\!\left[ -\left( \frac{c_-}{4\,c_s^2\,y_b'} \right) y^2 \right].
\end{equation}
Note that the variance of the Gaussian distribution obtained here from the analytical solution of non-conformal Landau hydrodynamics,
\begin{equation}\label{gaus_wid}
\sigma^2 = \frac{2\,c_s^2\,y_b'}{1-c_s^2},
\end{equation}
is different from those obtained previously by other authors \cite{Carruthers:1973ws, Shuryak:1972zq, Zhirov:1975qu, Gazdzicki:2010iv, Gao:2015mha} but agree in the conformal limit. Moreover, in these works, the authors have used $y_b'=\ln[\sqrt{s_{NN}}/2m_p]$ following Landau's original description \cite{Landau:1953gs} whereas we have employed an improved prescription $y_b'\equiv\frac{1}{2}ln[c_+/(4c_s^2)] + y_b$ following the arguments given in Ref.~\cite{Wong:2008ex}. However, within error bars, we did not obtain a clear signature for minima in $\sqrt{s_{NN}}$ dependence of $c_s^2$, even if we consider Landau's original prescription for $y_b'$ along with Eq.~\eqref{rap_gaus} for the fit.

%%%%%%%%%%%%%%%%%%%
%%%%%%%%%%%%%%%%%%%

In Fig.~\ref{spectra}, we have also shown the fit result for rapidity spectrum using the Gaussian distribution given in Eq.~\eqref{rap_gaus} (green dashed-dotted curves). We observe that the fit is very close to that obtained by using Eq.~\eqref{rap_sp}. Further, in Fig.~\ref{cs2}, we show a plot of squared speed of sound, extracted by fitting the pion rapidity spectra using Eq.~\eqref{rap_gaus} (green dashed-dotted line), over various collision energies (black dashed line). We see that the extracted value of $c_s^2$ is lower for higher $\sqrt{s_{NN}}$ but agree at lower collision energies. We found that using this Gaussian form, the fitted value of $c_s^2$ obtained for $\sqrt{s_{NN}}=200$~GeV matches with the conformal value of $1/3$. We see that even in this case, a minimum is not obtained in $\sqrt{s_{NN}}$ dependence of $c_s^2$. While viscosity could have affected the width of the Gaussian distribution, we have argued above that the effect is negligible. Therefore we claim that, for lower collision energies, rapidity spectra will provide a testing ground for determination of the correct value of $c_s^2$ and our solutions provide a framework for extracting this quantity. One should also keep in mind that the initial conditions may have significant effect on the evolution and should be correctly accounted for in such analysis.

%%%%%%%%%%%%%%%%%%%
%%%%%%%%%%%%%%%%%%%

At this juncture, we would like to reiterate that we have used a constant velocity of sound to derive the analytical expression for the evolution of energy density of the medium. We have employed this analytical expression to fit the observed rapidity spectra at different collision energies and extracted the numerical value of this constant sound velocity. However, lattice calculations predict temperature dependence of the sound velocity in the QCD medium \cite{Bazavov:2014pvz, Ding:2014kva}. Therefore one might view this value of extracted $c_s^2$ as the approximate time-averaged value for that particular collision energy \cite{Adare:2006ti, Csorgo:2018pxh}. This is in the same spirit as considering constant shear viscosity to entropy density ratio in hydrodynamic simulations where, in principle, one should consider temperature dependence. Our analytical expressions presented here will provide the platform to extract the average value of $c_s^2$ of the QCD medium formed in heavy-ion collisions.

%%%%%%%%%%%%%%%%%%Summary%%%%%%%%%%%%%%%%%%%%%%%%%%%%%%%%%%%%%%%%%%%%%%%%%

\emph{Summary and outlook}:
We have considered the relativistic Navier-Stokes equation for viscous evolution of the hot and dense matter formed in high energy heavy-ion collisions. We found analytical solution of the viscous evolution equation, for non-conformal system having constant speed of sound, with 1+1 dimensional Landau flow profile. We found that for Wong's modified prescription of Landau's freeze-out scenario, the viscous effects do not effect the rapidity spectrum of the produced particles. We employed the obtained solution to fit rapidity spectrum of observed pions in $\sqrt{s_{NN}}=$ 200, 17.3, 12.3, 8.76, 7.62, 6.27, 4.29, 3.83, 3.28 and 2.63 GeV collision energies. We showed that the obtained solutions of Landau flow with non-conformal equation of state leads to a better agreement with the experimental data compared to the conformal Landau flow solution. We also showed that the value of speed of sound, obtained by fitting the experimental results, show a monotonic decrease with decreasing collision energy as opposed to previous predictions of a minima.

The importance of the analysis performed here is its implications in extracting the value of $c_s^2$ of QCD medium formed in heavy-ion collisions by analyzing the rapidity spectrum of produced particles. More importantly, it is now well understood that hydrodynamic modeling of relativistic nuclear collisions at energies of order $10$ GeV requires treatment of non-equilibrium effects + breaking of boost invariance + non-zero baryon density. In this article, we have attempted to address the first two, i.e., treatment of non-equilibrium effects + breaking of boost invariance, within an analytical framework. Although we have performed the present calculations at vanishing baryon density, nevertheless, we venture to claim that the rapidity spectra will provide a testing ground for determination of the correct equation of state. We would like to emphasize that a simple analytical expression is always useful to understand the qualitative behavior of the physical system and helps in building intuition for the dynamics. At the very least, one can always treat an analytical solution of hydrodynamic equations as a benchmark to calibrate numerical codes.

Looking forward, it is of course important to obtain the solution of relativistic hydrodynamics and perform the analysis of rapidity spectra in presence of finite baryon chemical potential. Unlike Landau's original prescription of freeze-out hypersurface, which is also employed in the present work with Wong's modification, one needs to consider a finite temperature freeze-out in order to accurately estimate the viscous corrections to rapidity spectra. We leave these problems for future work.

%%%%%%%%%%%%%%%%%%Acknowledgments%%%%%%%%%%%%%%%%%%%%%%%%%%%%%%%%%%%%%%%%%

\medskip
\begin{acknowledgments}
\emph{Acknowledgments}:
We thank Samapan Bhadury, Sandeep Chatterjee and Najmul Haque for useful discussions. D.B. acknowledges support from UGC and thanks NISER for kind hospitality. A.J. is supported in part by the DST-INSPIRE faculty award under Grant No. DST/INSPIRE/04/2017/000038. S.R. is supported in part by the SERB Early Career Research Award under Grant No. ECR/2017/001354.
\end{acknowledgments}

%%%%%%%%%%%%%%%%%%%%%%%%%%%%%%%%%%%%%%%%%%%%%%%%%%%%%%%%%%%%%%%%%%%%%%%%%%


\begin{thebibliography}{99}

%\cite{Lee:1974ma}
\bibitem{Lee:1974ma} 
  T.~D.~Lee and G.~C.~Wick,
  %``Vacuum Stability and Vacuum Excitation in a Spin 0 Field Theory,''
  Phys.\ Rev.\ D {\bf 9}, 2291 (1974).
  %doi:10.1103/PhysRevD.9.2291
  %%CITATION = doi:10.1103/PhysRevD.9.2291;%%

%\cite{Collins:1974ky}
\bibitem{Collins:1974ky} 
  J.~C.~Collins and M.~J.~Perry,
  %``Superdense Matter: Neutrons Or Asymptotically Free Quarks?,''
  Phys.\ Rev.\ Lett.\  {\bf 34}, 1353 (1975).
  %%CITATION = PRLTA,34,1353;%%

%\cite{Itoh:1970uw}
\bibitem{Itoh:1970uw} 
  N.~Itoh,
  %``Hydrostatic Equilibrium of Hypothetical Quark Stars,''
  Prog.\ Theor.\ Phys.\  {\bf 44}, 291 (1970).
  %doi:10.1143/PTP.44.291
  %%CITATION = doi:10.1143/PTP.44.291;%%

%\cite{Romatschke:2007mq}
\bibitem{Romatschke:2007mq} 
  P.~Romatschke and U.~Romatschke,
  %``Viscosity Information from Relativistic Nuclear Collisions: How Perfect is the Fluid Observed at RHIC?,''
  Phys.\ Rev.\ Lett.\  {\bf 99}, 172301 (2007).
  %doi:10.1103/PhysRevLett.99.172301
  %[arXiv:0706.1522 [nucl-th]].
  %%CITATION = doi:10.1103/PhysRevLett.99.172301;%%

%\cite{Luzum:2008cw}
\bibitem{Luzum:2008cw} 
  M.~Luzum and P.~Romatschke,
  %``Conformal Relativistic Viscous Hydrodynamics: Applications to RHIC results at s(NN)**(1/2) = 200-GeV,''
  Phys.\ Rev.\ C {\bf 78}, 034915 (2008)
  Erratum: [Phys.\ Rev.\ C {\bf 79}, 039903 (2009)].
  %doi:10.1103/PhysRevC.78.034915, 10.1103/PhysRevC.79.039903
  %[arXiv:0804.4015 [nucl-th]].
  %%CITATION = doi:10.1103/PhysRevC.78.034915, 10.1103/PhysRevC.79.039903;%%

%\cite{Luzum:2009sb}
\bibitem{Luzum:2009sb} 
  M.~Luzum and P.~Romatschke,
  %``Viscous Hydrodynamic Predictions for Nuclear Collisions at the LHC,''
  Phys.\ Rev.\ Lett.\  {\bf 103}, 262302 (2009).
  %doi:10.1103/PhysRevLett.103.262302
  %[arXiv:0901.4588 [nucl-th]].
  %%CITATION = doi:10.1103/PhysRevLett.103.262302;%%

%\cite{Song:2010mg}
\bibitem{Song:2010mg} 
  H.~Song, S.~A.~Bass, U.~Heinz, T.~Hirano and C.~Shen,
  %``200 A GeV Au+Au collisions serve a nearly perfect quark-gluon liquid,''
  Phys.\ Rev.\ Lett.\  {\bf 106}, 192301 (2011)
  Erratum: [Phys.\ Rev.\ Lett.\  {\bf 109}, 139904 (2012)].
  %doi:10.1103/PhysRevLett.106.192301, 10.1103/PhysRevLett.109.139904
  %[arXiv:1011.2783 [nucl-th]].
  %%CITATION = doi:10.1103/PhysRevLett.106.192301, 10.1103/PhysRevLett.109.139904;%%

%\cite{Luzum:2010ag}
\bibitem{Luzum:2010ag} 
  M.~Luzum,
  %``Elliptic flow at energies available at the CERN Large Hadron Collider: Comparing heavy-ion data to viscous hydrodynamic predictions,''
  Phys.\ Rev.\ C {\bf 83}, 044911 (2011).
  %doi:10.1103/PhysRevC.83.044911
  %[arXiv:1011.5173 [nucl-th]].
  %%CITATION = doi:10.1103/PhysRevC.83.044911;%%
  
%\cite{Schenke:2011tv}
\bibitem{Schenke:2011tv} 
  B.~Schenke, S.~Jeon and C.~Gale,
  %``Anisotropic flow in $\sqrt{s}=2.76$ TeV Pb+Pb collisions at the LHC,''
  Phys.\ Lett.\ B {\bf 702}, 59 (2011).
  %doi:10.1016/j.physletb.2011.06.065
  %[arXiv:1102.0575 [hep-ph]].
  %%CITATION = doi:10.1016/j.physletb.2011.06.065;%%

%\cite{Gale:2012rq}
\bibitem{Gale:2012rq} 
  C.~Gale, S.~Jeon, B.~Schenke, P.~Tribedy and R.~Venugopalan,
  %``Event-by-event anisotropic flow in heavy-ion collisions from combined Yang-Mills and viscous fluid dynamics,''
  Phys.\ Rev.\ Lett.\  {\bf 110}, 012302 (2013).
  %doi:10.1103/PhysRevLett.110.012302
  %[arXiv:1209.6330 [nucl-th]].
  %%CITATION = doi:10.1103/PhysRevLett.110.012302;%%

%\cite{Bhalerao:2015iya}
\bibitem{Bhalerao:2015iya} 
  R.~S.~Bhalerao, A.~Jaiswal and S.~Pal,
  %``Collective flow in event-by-event partonic transport plus hydrodynamics hybrid approach,''
  Phys.\ Rev.\ C {\bf 92}, 014903 (2015).
  %doi:10.1103/PhysRevC.92.014903
  %[arXiv:1503.03862 [nucl-th]].
  %%CITATION = doi:10.1103/PhysRevC.92.014903;%%

%\cite{Jaiswal:2016hex}
\bibitem{Jaiswal:2016hex} 
  A.~Jaiswal and V.~Roy,
  %``Relativistic hydrodynamics in heavy-ion collisions: general aspects and recent developments,''
  Adv.\ High Energy Phys.\  {\bf 2016}, 9623034 (2016).
  %doi:10.1155/2016/9623034
  %[arXiv:1605.08694 [nucl-th]].
  %%CITATION = doi:10.1155/2016/9623034;%%

%\cite{Landau:1953gs}
\bibitem{Landau:1953gs} 
  L.~D.~Landau,
  %``On the multiparticle production in high-energy collisions,''
  Izv.\ Akad.\ Nauk Ser.\ Fiz.\  {\bf 17}, 51 (1953).
  %%CITATION = IANFA,17,51;%%
  
%\cite{Hwa:1974gn}
\bibitem{Hwa:1974gn} 
  R.~C.~Hwa,
  %``Statistical Description of Hadron Constituents as a Basis for the Fluid Model of High-Energy Collisions,''
  Phys.\ Rev.\ D {\bf 10}, 2260 (1974).
  %doi:10.1103/PhysRevD.10.2260
  %%CITATION = doi:10.1103/PhysRevD.10.2260;%%
  
%\cite{Bjorken:1982qr}
\bibitem{Bjorken:1982qr} 
  J.~D.~Bjorken,
  %``Highly Relativistic Nucleus-Nucleus Collisions: The Central Rapidity Region,''
  Phys.\ Rev.\ D {\bf 27}, 140 (1983).
  %doi:10.1103/PhysRevD.27.140
  %%CITATION = doi:10.1103/PhysRevD.27.140;%%
  
%\cite{Murray:2004gh}
\bibitem{Murray:2004gh} 
  M.~J.~Murray [BRAHMS Collaboration],
  %``Scanning the phases of QCD with BRAHMS,''
  J.\ Phys.\ G {\bf 30}, S667 (2004).
  %doi:10.1088/0954-3899/30/8/004
  %[nucl-ex/0404007].
  %%CITATION = doi:10.1088/0954-3899/30/8/004;%%
  
%\cite{Bearden:2004yx}
\bibitem{Bearden:2004yx} 
  I.~G.~Bearden {\it et al.} [BRAHMS Collaboration],
  %``Charged meson rapidity distributions in central Au+Au collisions at s(NN)**(1/2) = 200-GeV,''
  Phys.\ Rev.\ Lett.\  {\bf 94}, 162301 (2005).
  %doi:10.1103/PhysRevLett.94.162301
  %[nucl-ex/0403050].
  %%CITATION = doi:10.1103/PhysRevLett.94.162301;%%
  
%\cite{Murray:2007cy}
\bibitem{Murray:2007cy} 
  M.~Murray [BRAHMS Collaboration],
  %``Flavor dynamics,''
  J.\ Phys.\ G {\bf 35}, 044015 (2008).
  %doi:10.1088/0954-3899/35/4/044015
  %[arXiv:0710.4576 [nucl-ex]].
  %%CITATION = doi:10.1088/0954-3899/35/4/044015;%%
  
%\cite{Steinberg:2004wx}
\bibitem{Steinberg:2004wx} 
  P.~Steinberg,
  %``Bulk dynamics in heavy ion collisions,''
  Nucl.\ Phys.\ A {\bf 752}, 423 (2005).
  %doi:10.1016/j.nuclphysa.2005.02.139
  %[nucl-ex/0412009].
  %%CITATION = doi:10.1016/j.nuclphysa.2005.02.139;%%
  
%\cite{Steinberg:2007iv}
\bibitem{Steinberg:2007iv} 
  P.~Steinberg,
  %``Entropy Production at High Energy and mu(B),''
  PoS CPOD {\bf 2006}, 036 (2006).
  %doi:10.22323/1.029.0036
  %[nucl-ex/0702019 [NUCL-EX]].
  %%CITATION = doi:10.22323/1.029.0036;%%
  
%\cite{Wong:2008ex}
\bibitem{Wong:2008ex} 
  C.~Y.~Wong,
  %``Landau Hydrodynamics Revisited,''
  Phys.\ Rev.\ C {\bf 78}, 054902 (2008).
  %doi:10.1103/PhysRevC.78.054902
  %[arXiv:0808.1294 [hep-ph]].
  %%CITATION = doi:10.1103/PhysRevC.78.054902;%% 

%\cite{Belenkij:1956cd}
\bibitem{Belenkij:1956cd} 
  S.~Z.~Belenkij and L.~D.~Landau,
  %``Hydrodynamic theory of multiple production of particles,''
  Nuovo Cim.\ Suppl.\  {\bf 3S10}, 15 (1956)
  [Usp.\ Fiz.\ Nauk {\bf 56}, 309 (1955)].
  %doi:10.1007/BF02745507
  %%CITATION = doi:10.1007/BF02745507;%%

%\cite{Khalatnikov:1954}
\bibitem{Khalatnikov:1954} 
  I.~M.~Khalatnikov, 
  %``Some questions of relativistic hydrodynamics,"
  Zh. Eksp. Teor. Fiz. {\bf 27}, 529 (1954).

%\cite{Belenkij:1956}
\bibitem{Belenkij:1956}
  S.~Z.~Belenkij and G.~A.~Milekhin,
  Zh. Eksp. Teor. Fiz. {\bf 29}, 20 (1956) 
  [Sov. Phys. JETP {\bf 2}, 14 (1956)].
    
%\cite{Rosental:1957}
\bibitem{Rosental:1957}
  I.~L.~Rosental, 
  Zh. Eksp. Teor. Fiz. {\bf 31}, 278 (1957) 
  [Sov. Phys. JETP {\bf 4}, 217 (1959)].
  
%\cite{Milekhin:1958a}
\bibitem{Milekhin:1958a}
  G.~A.~Milekhin,
  Zh. Eksp. Teor. Fiz. {\bf 35}, 978 (1958) 
  [Sov. Phys. JETP {\bf 8}, 682 (1959)].
    
%\cite{Milekhin:1958b}
\bibitem{Milekhin:1958b}
  G.~A.~Milekhin, 
  Zh. Eksp. Teor. Fiz. {\bf 35}, 1185 (1958) 
  [Sov. Phys. JETP {\bf 8}, 829 (1959)].
  
%\cite{Amai:1957}
\bibitem{Amai:1957}
  S.~Amai, H.~Fukuda, C.~Iso, and M.~Sato, 
  Prog. Theo. Phys. {\bf 17}, 241 (1957).

%\cite{Carruthers:1973ws}
\bibitem{Carruthers:1973ws} 
  P.~Carruthers and M.~Doung-van,
  %``Rapidity and angular distributions of charged secondaries according to the hydrodynamical model of particle production,''
  Phys.\ Rev.\ D {\bf 8}, 859 (1973).
  %doi:10.1103/PhysRevD.8.859
  %%CITATION = doi:10.1103/PhysRevD.8.859;%%

%\cite{Cooper:1974mv}
\bibitem{Cooper:1974mv} 
  F.~Cooper and G.~Frye,
  %``Comment on the Single Particle Distribution in the Hydrodynamic and Statistical Thermodynamic Models of Multiparticle Production,''
  Phys.\ Rev.\ D {\bf 10}, 186 (1974).
  %doi:10.1103/PhysRevD.10.186
  %%CITATION = doi:10.1103/PhysRevD.10.186;%%

%\cite{Cooper:1974qi}
\bibitem{Cooper:1974qi} 
  F.~Cooper, G.~Frye and E.~Schonberg,
  %``Landau's Hydrodynamic Model of Particle Production and electron Positron Annihilation Into Hadrons,''
  Phys.\ Rev.\ D {\bf 11}, 192 (1975).
  %doi:10.1103/PhysRevD.11.192
  %%CITATION = doi:10.1103/PhysRevD.11.192;%%

%\cite{Chadha:1974qs}
\bibitem{Chadha:1974qs} 
  S.~Chadha, C.~S.~Lam and Y.~C.~Leung,
  %``On Proton Proton Collisions in the Hydrodynamic Theory,''
  Phys.\ Rev.\ D {\bf 10}, 2817 (1974).
  %doi:10.1103/PhysRevD.10.2817
  %%CITATION = doi:10.1103/PhysRevD.10.2817;%%
  
%\cite{Srivastava:1992xb}
\bibitem{Srivastava:1992xb} 
  D.~K.~Srivastava, J.~Alam and B.~Sinha,
  %``Rapidity distribution of secondaries in ultrarelativistic heavy ion collisions using Landau's hydrodynamic model,''
  Phys.\ Lett.\ B {\bf 296}, 11 (1992).
  %doi:10.1016/0370-2693(92)90796-7
  %%CITATION = doi:10.1016/0370-2693(92)90796-7;%%

%\cite{Srivastava:1992cg}
\bibitem{Srivastava:1992cg} 
  D.~K.~Srivastava, J.~e.~Alam, S.~Chakrabarty, B.~Sinha and S.~Raha,
  %``Hydrodynamics of ultrarelativistic heavy ion collisions: Considerations of boost noninvariance and stopping,''
  Annals Phys.\  {\bf 228}, 104 (1993).
  %doi:10.1006/aphy.1993.1089
  %%CITATION = doi:10.1006/aphy.1993.1089;%%

%\cite{Srivastava:1992gh}
\bibitem{Srivastava:1992gh} 
  D.~K.~Srivastava, J.~Alam, S.~Chakrabarty, S.~Raha and B.~Sinha,
  %``Boost noninvariant hydrodynamics in ultrarelativistic heavy ion collisions,''
  Phys.\ Lett.\ B {\bf 278}, 225 (1992).
  %doi:10.1016/0370-2693(92)90185-7
  %%CITATION = doi:10.1016/0370-2693(92)90185-7;%%
  
%\cite{Mohanty:2003va}
\bibitem{Mohanty:2003va} 
  B.~Mohanty and J.~e.~Alam,
  %``Velocity of sound in relativistic heavy ion collisions,''
  Phys.\ Rev.\ C {\bf 68}, 064903 (2003).
  %doi:10.1103/PhysRevC.68.064903
  %[nucl-th/0301086].
  %%CITATION = doi:10.1103/PhysRevC.68.064903;%%
  
%\cite{Hama:2004rr}
\bibitem{Hama:2004rr} 
  Y.~Hama, T.~Kodama and O.~Socolowski, Jr.,
  %``Topics on hydrodynamic model of nucleus-nucleus collisions,''
  Braz.\ J.\ Phys.\  {\bf 35}, 24 (2005).
  %doi:10.1590/S0103-97332005000100003
  %[hep-ph/0407264].
  %%CITATION = doi:10.1590/S0103-97332005000100003;%%

%\cite{Aguiar:2000hw}
\bibitem{Aguiar:2000hw} 
  C.~E.~Aguiar, T.~Kodama, T.~Osada and Y.~Hama,
  %``Smoothed particle hydrodynamics for relativistic heavy ion collisions,''
  J.\ Phys.\ G {\bf 27}, 75 (2001).
  %doi:10.1088/0954-3899/27/1/306
  %[hep-ph/0006239].
  %%CITATION = doi:10.1088/0954-3899/27/1/306;%%
  
%\cite{Pratt:2008jj}
\bibitem{Pratt:2008jj} 
  S.~Pratt,
  %``A Co-moving Coordinate System for Relativistic Hydrodynamics,''
  Phys.\ Rev.\ C {\bf 75}, 024907 (2007).
  %doi:10.1103/PhysRevC.75.024907
  %[nucl-th/0612010].
  %%CITATION = doi:10.1103/PhysRevC.75.024907;%%

%\cite{Bialas:2007iu}
\bibitem{Bialas:2007iu} 
  A.~Bialas, R.~A.~Janik and R.~B.~Peschanski,
  %``Unified description of Bjorken and Landau 1+1 hydrodynamics,''
  Phys.\ Rev.\ C {\bf 76}, 054901 (2007).
  %doi:10.1103/PhysRevC.76.054901
  %[arXiv:0706.2108 [nucl-th]].
  %%CITATION = doi:10.1103/PhysRevC.76.054901;%%  
   
%\cite{Csorgo:2006ax}
\bibitem{Csorgo:2006ax} 
  T.~Csorgo, M.~I.~Nagy and M.~Csanad,
  %``A New family of simple solutions of perfect fluid hydrodynamics,''
  Phys.\ Lett.\ B {\bf 663}, 306 (2008).
  %doi:10.1016/j.physletb.2008.04.298
  %[nucl-th/0605070].
  %%CITATION = doi:10.1016/j.physletb.2008.04.038;%%
  
%\cite{Beuf:2008vd}
\bibitem{Beuf:2008vd} 
  G.~Beuf, R.~Peschanski and E.~N.~Saridakis,
  %``Entropy flow of a perfect fluid in (1+1) hydrodynamics,''
  Phys.\ Rev.\ C {\bf 78}, 064909 (2008).
  %doi:10.1103/PhysRevC.78.064909
  %[arXiv:0808.1073 [nucl-th]].
  %%CITATION = doi:10.1103/PhysRevC.78.064909;%%
  
%\cite{Osada:2008cn}
\bibitem{Osada:2008cn} 
  T.~Osada and G.~Wilk,
  %``Nonextensive perfect hydrodynamics: A Model of dissipative relativistic hydrodynamics?,''
  Central Eur.\ J.\ Phys.\  {\bf 7}, 432 (2009).
  %doi:10.2478/s11534-008-0163-5
  %[arXiv:0810.3089 [hep-ph]].
  %%CITATION = doi:10.2478/s11534-008-0163-5;%%
  
%\cite{Hoang:1973uz}
\bibitem{Hoang:1973uz} 
  T.~F.~Hoang,
  %``Viscosity effect in landau's hydrodynamical model of particle production,''
  Nuovo Cim.\ A {\bf 17}, 625 (1973).
  %doi:10.1007/BF02786838
  %%CITATION = doi:10.1007/BF02786838;%%  
  
%\cite{Chaichian:1976nt}
\bibitem{Chaichian:1976nt} 
  M.~Chaichian and E.~Suhonen,
  %``Corrections to the Landau Hydrodynamical Model of Multiparticle Production Due to Viscosity,''
  Nucl.\ Phys.\ B {\bf 127}, 461 (1977).
  %doi:10.1016/0550-3213(77)90451-5
  %%CITATION = doi:10.1016/0550-3213(77)90451-5;%%

%\cite{Hoang:1977wf}
\bibitem{Hoang:1977wf} 
  T.~F.~Hoang and K.~K.~Phua,
  %``Viscosity Effect in Landau's Hydrodynamical Model,''
  Z.\ Phys.\ C {\bf 2}, 295 (1979).
  %doi:10.1007/BF01545889
  %%CITATION = doi:10.1007/BF01545889;%%

%\cite{Landau_book} 
\bibitem{Landau_book} 
  L.~D.~Landau and E.~M.~Lifshitz, 
  {\it Fluid Mechanics}
  (Butterworth-Heinemann, Oxford, 1987).
  
%\cite{Hosoya:1983xm}
\bibitem{Hosoya:1983xm} 
  A.~Hosoya and K.~Kajantie,
  %``Transport Coefficients of QCD Matter,''
  Nucl.\ Phys.\ B {\bf 250}, 666 (1985).
  %doi:10.1016/0550-3213(85)90499-7
  %%CITATION = doi:10.1016/0550-3213(85)90499-7;%%  
  
%\cite{Kouno:1989ps}
\bibitem{Kouno:1989ps} 
  H.~Kouno, M.~Maruyama, F.~Takagi and K.~Saito,
  %``Relativistic Hydrodynamics of Quark - Gluon Plasma and Stability of Scaling Solutions,''
  Phys.\ Rev.\ D {\bf 41}, 2903 (1990).
  %doi:10.1103/PhysRevD.41.2903
  %%CITATION = doi:10.1103/PhysRevD.41.2903;%%
  
%\cite{Teaney:2003kp}
\bibitem{Teaney:2003kp} 
  D.~Teaney,
  %``The Effects of viscosity on spectra, elliptic flow, and HBT radii,''
  Phys.\ Rev.\ C {\bf 68}, 034913 (2003).
  %doi:10.1103/PhysRevC.68.034913
  %[nucl-th/0301099].
  %%CITATION = doi:10.1103/PhysRevC.68.034913;%%
  
%\cite{Song:2009rh}
\bibitem{Song:2009rh} 
  H.~Song and U.~W.~Heinz,
  %``Interplay of shear and bulk viscosity in generating flow in heavy-ion collisions,''
  Phys.\ Rev.\ C {\bf 81}, 024905 (2010).
  %doi:10.1103/PhysRevC.81.024905
  %[arXiv:0909.1549 [nucl-th]].
  %%CITATION = doi:10.1103/PhysRevC.81.024905;%%
  
%\cite{Klay:2003zf}
\bibitem{Klay:2003zf} 
  J.~L.~Klay {\it et al.} [E-0895 Collaboration],
  %``Charged pion production in 2 to 8 agev central au+au collisions,''
  Phys.\ Rev.\ C {\bf 68}, 054905 (2003).
  %doi:10.1103/PhysRevC.68.054905
  %[nucl-ex/0306033].
  %%CITATION = doi:10.1103/PhysRevC.68.054905;%%
  
%\cite{Alt:2007aa}
\bibitem{Alt:2007aa} 
  C.~Alt {\it et al.} [NA49 Collaboration],
  %``Pion and kaon production in central Pb + Pb collisions at 20-A and 30-A-GeV: Evidence for the onset of deconfinement,''
  Phys.\ Rev.\ C {\bf 77}, 024903 (2008).
  %doi:10.1103/PhysRevC.77.024903
  %[arXiv:0710.0118 [nucl-ex]].
  %%CITATION = doi:10.1103/PhysRevC.77.024903;%%
  
%\cite{Afanasiev:2002mx}
\bibitem{Afanasiev:2002mx} 
  S.~V.~Afanasiev {\it et al.} [NA49 Collaboration],
  %``Energy dependence of pion and kaon production in central Pb + Pb collisions,''
  Phys.\ Rev.\ C {\bf 66}, 054902 (2002).
  %doi:10.1103/PhysRevC.66.054902
  %[nucl-ex/0205002].
  %%CITATION = doi:10.1103/PhysRevC.66.054902;%%

%\cite{Gazdzicki:2010iv}  
\bibitem{Gazdzicki:2010iv} 
  M.~Gazdzicki, M.~Gorenstein and P.~Seyboth,
  %``Onset of deconfinement in nucleus-nucleus collisions: Review for pedestrians and experts,''
  Acta Phys.\ Polon.\ B {\bf 42}, 307 (2011).
  %doi:10.5506/APhysPolB.42.307
  %[arXiv:1006.1765 [hep-ph]].
  %%CITATION = doi:10.5506/APhysPolB.42.307;%%  

%\cite{Gao:2015mha}
\bibitem{Gao:2015mha} 
  L.~N.~Gao and F.~H.~Liu,
  %``On Distributions of Emission Sources and Speed of Sound in Proton-proton (Proton-antiproton) Collisions,''
  Adv.\ High Energy Phys.\  {\bf 2015}, 641906 (2015).
  %doi:10.1155/2015/641906
  %[arXiv:1509.09034 [nucl-th]].
  %%CITATION = doi:10.1155/2015/641906;%%
  
%\cite{Bazavov:2014pvz}
\bibitem{Bazavov:2014pvz} 
  A.~Bazavov {\it et al.} [HotQCD Collaboration],
  %``Equation of state in ( 2+1 )-flavor QCD,''
  Phys.\ Rev.\ D {\bf 90}, 094503 (2014).
  %doi:10.1103/PhysRevD.90.094503
  %[arXiv:1407.6387 [hep-lat]].
  %%CITATION = doi:10.1103/PhysRevD.90.094503;%%  

%\cite{Ding:2014kva}
\bibitem{Ding:2014kva} 
  H.~T.~Ding,
  %``Recent lattice QCD results and phase diagram of strongly interacting matter,''
  Nucl.\ Phys.\ A {\bf 931}, 52 (2014).
  %doi:10.1016/j.nuclphysa.2014.09.053
  %[arXiv:1408.5236 [hep-lat]].
  %%CITATION = doi:10.1016/j.nuclphysa.2014.09.053;%%
  
%\cite{Shuryak:1972zq}
\bibitem{Shuryak:1972zq} 
  E.~V.~Shuryak,
  %``Multiparticle production in high energy particle collisions.,''
  Yad.\ Fiz.\  {\bf 16}, 395 (1972).
  %%CITATION = YAFIA,16,395;%%

%\cite{Zhirov:1975qu}
\bibitem{Zhirov:1975qu} 
  O.~V.~Zhirov and E.~V.~Shuryak,
  %``Multiple Production of Hadrons and Predictions of the Landau Theory,''
  Yad.\ Fiz.\  {\bf 21}, 861 (1975).
  %%CITATION = YAFIA,21,861;%%
  
%\cite{Adare:2006ti}
\bibitem{Adare:2006ti} 
  A.~Adare {\it et al.} [PHENIX Collaboration],
  %``Scaling properties of azimuthal anisotropy in Au+Au and Cu+Cu collisions at s(NN) = 200-GeV,''
  Phys.\ Rev.\ Lett.\  {\bf 98}, 162301 (2007).
  %doi:10.1103/PhysRevLett.98.162301
  %[nucl-ex/0608033].
  %%CITATION = doi:10.1103/PhysRevLett.98.162301;%%

%\cite{Csorgo:2018pxh}
\bibitem{Csorgo:2018pxh} 
  T.~Csörgő, G.~Kasza, M.~Csanád and Z.~Jiang,
  %``New exact solutions of relativistic hydrodynamics for longitudinally expanding fireballs,''
  Universe {\bf 4}, no. 6, 69 (2018).
  %doi:10.3390/universe4060069
  %[arXiv:1805.01427 [nucl-th]].
  %%CITATION = doi:10.3390/universe4060069;%%   
  
\end{thebibliography}
\end{document}